\documentclass{PoS}

\usepackage{graphicx}
\usepackage{latexsym}
\usepackage{amssymb,amsmath}
\usepackage{epsfig}

\def\bb{\begin{equation*}}
\def\eb{\end{equation*}}
\def\beb{\begin{eqnarray*}}
\def\eeb{\end{eqnarray*}}
\def\be{\begin{equation}}
\def\ee{\end{equation}}
\def\bea{\begin{eqnarray}}
\def\eea{\end{eqnarray}}
\def\re{\rm e}
\def\ri{\rm i}

\newcommand{\starco}[2]{\left[ #1\stackrel{\star}{,}#2\right] }
\newcommand{\staraco}[2]{\left\{ #1\stackrel{\star}{,}#2\right\} }

\title{Noncommutative gauge theory and renormalisability}

\ShortTitle{NC gauge theories}

\author{\speaker{Michael Wohlgenannt}\\
        University of Vienna, Faculty of Physics\\
        Boltzmanngasse 5, A-1090 Vienna, Austria\\
        E-mail: \email{michael.wohlgenannt@univie.ac.at}}

\abstract{We review two different noncommutative gauge models generalizing approaches which lead to renormalizable 
scalar quantum field theories. One of them implements the crucial IR damping of the gauge field propagator in the 
so-called ``soft breaking'' part. We discuss one-loop renormalisability.}

\FullConference{Corfu Summer Institute on Elementary Particles and Physics - Workshop on NonCommutative Field Theory and Gravity,\\
September 8-12, 2010\\ Corfu Greece}

\begin{document}


\section{Introduction}

There are various motivations for studying noncommutative geometries. They range from general considerations 
in Quantum Field Theory (QFT) \cite{Schroedinger:1934aa,Heisenberg:1938aa} and (Quantum) Gravity 
\cite{Doplicher:1994tu,Garay:1995en} to String Theory and Matrix Models \cite{Connes:1997cr,Seiberg:1999vs,Schomerus:1999ug} 
and purely mathematical considerations \cite{Connes:1986uh}. One of the first applications of noncommutative ideas was already 
within the realm of gauge theories, namely the Quantum Hall effect \cite{Bellissard:1986gt}. What is most remarkable in my eyes,
is the intimate connection between noncommutative gauge theory and gravity. This connection is not fully understood 
at present and studied from different points of view, see e.g. 
\cite{Szabo:2006wx,Yang:2006dk,Calmet:2006iz,Steinacker:2007dq,Chaichian:2007we,Balachandran:2007kv}
and references therein for a merely exemplary list of quotations.

In this note, we concentrate on models for noncommutative gauge theories, where the idea of renormalizability
will be a guiding principle. Furthermore, we consider canonically deformed 4D Euclidean space. The coordinates 
satisfy the following commutation relations
\be
[x^i\stackrel{\star}{,} x^j] = \textrm{i} \Theta^{ij}\,,
\ee
where $\Theta^{ij}=-\Theta^{ji}=\textrm{const}$, and the star product is given by the Moyal-Weyl product,
\be
f \star g\, (x) = \textrm{e}^{\frac{\textrm{i}}2 \Theta^{ij}\partial^x_i \partial^y_j}
f(x)\, g(y)\Big|_{y\to x}\,.
\ee

In the next section, we will discuss the so-called UV/IR mixing problem in the case of scalar field theory. 
It is a thread to renormalizability. Up to now, there are two different models which overcome this problem and
which are perturbatively renormalizable to all orders. Both are formulated on canonically deformed Euclidean space. 
In Section \ref{gauge}, we will attempt to generalize both approaches to noncommutative $U(1)$ gauge theory.
A brief summary and some concluding remarks follow in Section \ref{conclusions}. 


\section{UV/IR mixing in scalar theories}

The simplest approach to noncommutative $\phi^4$ theory is to take the commutative action and
replace the pointwise products by star products. Since the star product is not relevant for bilinear expressions,
only the selfinteraction term is modified, and we obtain
\be
\label{action1}
S = \int d^4x \, \left( \frac 12
\partial_\mu \phi\, \partial_\mu \phi + \frac{m^2}2 \phi^2 + \frac \lambda {4!} \phi\star \phi \star \phi\star \phi
\right)\,.
\ee
The above action determines the Feynman rules. The propagator is the same as in the commutative case, 
\be
G(p) = \frac1{p^2+m^2}\,,
\ee
while the vertex is decorated by momentum dependent phase factors:
\be
\Gamma (p_1,\dots,p_4) = - \lambda \delta^{(4)}(p_1+p_2+p_3+p_4) \textrm{e}^{-\textrm{i} \sum_{i<j} p_i \Theta p_j}\,.
\ee
As a consequence, new types of Feynman graphs occur: In addition to the ones known from commutative space, where no 
phases depending on internal loop momenta appear and showing              the usual UV divergences, so-called non-planar graphs 
come into the game which are regularized by phases depending on internal momenta. 
One-loop calculations have been performed explicitly \cite{Minwalla:1999px,Belov:2000,Matusis:2000jf,Micu:2000,Schweda:2002b} 
and hence the UV/IR mixing problem has been found: Due to the phases in the non-planar graphs, their UV sector is regularized 
on the one hand, but on the other hand this regularization implies divergences for small external momenta. 
For example, let us consider the two point tadpole graph. It is given by the expression
\be
\Pi(\Lambda,p) \propto \lambda \int d^4k \frac{2+\cos (k\tilde p)}{k^2+m^2} = \Pi^{UV}(\Lambda) + \Pi^{IR}(\Lambda,p)\,.
\ee
The planar contribution is as usual quadratically divergent in the UV cutoff $\Lambda$, i.e. $\Pi^{UV}\sim \Lambda^2$, and the 
non-planar part is regularized by the cosine,
\be
\label{ir-sing}
\Pi^{IR} \sim \frac 1{\tilde p^2}\,,
\ee
where $\tilde p_\mu = \Theta_{\mu\nu}p_\nu$.
The original UV divergence is not present, but reappears when $\tilde p\to 0$ re\-presenting a new kind of infrared 
divergence. Since both divergences are related to one another, one speaks of ``UV/IR mixing''. At one-loop level, this is no
problem though. It corresponds to a counter term
\be
\label{ct1}
\int d^4p\, \tilde \phi(p) \frac 1{\tilde p^2} \, \tilde \phi(-p)\,,
\ee
which is well behaved even in the limit $\tilde p\to 0$. But higher loop insertions then lead to a term of the form 
\be
\label{ct2}
\int d^4p\, \tilde \phi(p) \frac 1{(\tilde p^2)^n} \tilde \phi(-p)\,,
\ee
where $n$ is the number of insertions. Clearly, this term exhibits a serious IR singularity.
It is this mixing which renders the action \eqref{action1} non-renormalizable. Two different 
strategies to cure UV/IR mixing are known. Both modify the propagator by adding an additional 
term quadratic in the fields: An oscillator term (Section \ref{gw}) and a $1/\tilde p^2$-term 
(Section \ref{pp}), respectively.In what follows, we will briefly review those approaches.


\subsection{\label{gw}The scalar Grosse-Wulkenhaar model}

Adding an oscillator potential and after some awkward rewritting, the action \eqref{action1} becomes 
\cite{Grosse:2003nw,Grosse:2004yu} 
\bea 
\label{scalar-oscillator}
S & = & \int d^4x 
\left( \frac{1}{2} \phi \star \starco{\tilde x_\nu}{\starco{\tilde x^\nu}{\phi}} 
+ \frac{\Omega^2}{2} \phi \star \staraco{\tilde x_\nu}{\starco{\tilde x^\nu}{\phi}}  \right.
\\
&& + \left. \frac{\mu^2}{2} \phi \star \phi + \frac{\lambda}{4!} 
\phi\star \phi  \star \phi \star \phi\right) \,, 
\nonumber 
\eea 
where $\tilde x_\nu = \theta^{-1}_{\nu\alpha}x^\alpha$, and we have used 
$\ri \,\partial_\mu f = \starco{\tilde x_\mu}{f}$. This action is covariant, i.e.
\be
 S[\phi;\mu,\lambda,\Omega]\mapsto\Omega^2S[\phi;\frac{\mu}{\Omega},\frac{\lambda}{\Omega^2},\frac{1}{\Omega}]\,,
\ee 
under the so-called Langmann-Szabo duality transformation \cite{Langmann:2002cc} between position and momenta:
\be
\hat \phi(p) \longleftrightarrow \pi^2 \sqrt{|det \Theta|}\, \phi(x)\,, \qquad
p_\mu \longleftrightarrow 2 \tilde x_\mu\,,
\ee
where $\hat \phi(p_a)=\int d^4x_a\re^{(-1)^a \ri p_{a,\mu}x_{a,\mu}}\, \phi(x_a)$. The index $a$ is 
labelling the legs of vertex and propagator, resp. and defines the direction of the according momentum. 
This becomes a symmetry at $\Omega=1$.Due to oscillator term, the propagator is modified and an IR damping 
is implemented. The propagator is given by the Mehler kernel:
\be
\label{mehler}
K_M(p,q) = \frac{\omega^3}{8\pi^2} \int_0^\infty \frac{d\alpha}{\sinh^2\alpha}\,
\textrm{e}^{-\frac\omega 4 (p-q)^2 \coth\frac \alpha 2 - \frac \omega 4 (p+q)^2 \tanh \frac\alpha 2}\,,
\ee
where $\omega= \Theta/\Omega$. The IR damping is also responsible for a proper handling of the UV/IR mixing 
problem. The model is renormalisable to all orders in perturbation theory. The propagator depends on two 
momenta, an incoming and outgoing momentum, since the explicit $x$-dependence of the action breaks translation 
invariance. Therefore, also momentum conservation is broken.
Remarkably, the oscillator term can be interpreted as coupling of the scalar field to the 
curvature of some specific noncommutative background \cite{Buric:2009ss}.


\subsection{\label{pp}$1/p^2$ model}

In the second approach, a non-local term is added to the action \eqref{action1}. In momentum space, it reads
\cite{Gurau:2008vd}
\be
\label{ct1a}
S_{nl} = \int d^4p\, \frac a2 \tilde\phi(p) \frac 1{\tilde p^2}\, \tilde \phi (-p)\,.
\ee
This is exactly the counter term \eqref{ct1} we have discussed before. The resulting action is translation 
invariant, and thus momentum conservation holds. The term \eqref{ct1a} implements IR damping for the propagator, 
i.e. $G(p)\to 0$, for $p\to 0$. The modified propagator has the form
\be
\label{propagator-pp}
G(p) = \frac 1{p^2 + m^2 + \frac {a^2}{p^2}}\,.
\ee
The damping effect of the propagator becomes obvious when one considers higher loop orders. An $n$-fold insertion 
of the divergent one-loop result \eqref{ir-sing} into a single large loop can be written as
\be
\label{eq:canon:p2inv_gurau_n-loop-int}
\Pi^{n {\rm np-ins.}}(p) \approx \lambda^2 \int d^4k
\, \frac{\re^{\ri k \tilde p}}{\left(\tilde k^2\right)^n\left[k^2+m^2+\frac{a'^2}{k^2}\right]^{n+1}}\,,
\ee
neglecting any effects due to recursive renormalization and approximating the insertions of irregular single loops 
by the most divergent (quadratic) IR divergence. For the model \eqref{action1}, i.e. $a=0$, the integrand is
proportional to $(k^2)^{-n}$, for $k^2\to 0$, as we have already mentioned. But $a\neq 0$ implies that the integrand behaves like
\be
\label{eq:canon:p2inv_gurau_n-loop-int-approx}
\frac 1{\left(\tilde k^2\right)^n\left[\frac{a'^2}{k^2}\right]^{n+1}}=\frac{\tilde k^2}{\left(a'^2\right)^{n+1}}\,,
\ee
which is independent of the loop order $n$. Using multiscale analysis, the perturbative renormalisability of this
model to all orders could be shown \cite{Gurau:2008vd}.


\section{\label{gauge}Noncommutative gauge theory}

The aim of this section is to generalize the approaches discussed above to 
noncommutative $U(1)$ gauge theory. They are good candidates for rernomalizable models.
As we will see, UV/IR mixing also occurs in the case of noncommutative gauge theory, and so far, no model could be shown to be renormalisable.


\subsection{Oscillator approach}

As a first step, a BRST invariant action including an oscillator term has been proposed in \cite{Blaschke:2007vc}:
\be
\label{oscillator-action-1}
S = \int d^4 x \bigg( \frac14 F_{\mu\nu}\star F^{\mu\nu} 
+ s(\bar c\star \partial_\mu A_\mu) - \frac12 B^2 + \frac{\Omega^2}8 
s(\tilde c_\mu \star \mathcal C_\mu) \bigg) \,,
\ee
where $\mathcal C_\mu$ contains the crucial new terms:
\be
\mathcal C_\mu = \staraco{\staraco{\tilde x_\mu}{A_\nu}}{A_\nu} + 
\starco{\staraco{\tilde x_\mu}{\bar c}}{c} + \starco{\bar c}{\staraco{\tilde x_\mu}{c}}\,,
\ee
and $\tilde c_\mu$ is a new parameter which also transforms under BRST. The noncommutative field strength is given by
$F_{\mu\nu} = \partial_\mu A_\nu - \partial_\nu A_\mu - \ri\starco{A_\mu}{A_\nu}$.
Summing up, the action \eqref{oscillator-action-1} is invariant under the following BRST transformation:
\bea
sA_\mu & = & D_\mu c, \quad s\bar c = B, \qquad sc = ig c\star c,\\
\nonumber
sB & = & 0, \qquad s\tilde c_\mu = \tilde x_\mu\,.
\eea
The above set of transformations is nilpotent. The propagator of the gauge field is given by Mehler kernel 
\eqref{mehler}. One-loop calculations have been preformed in \cite{Blaschke:2009aw}. A power counting formula 
has been obtained and the corrections to the vertex functions have been computed. Remarkably, the one-point 
tadpole is UV-divergent. Therefore, the action \eqref{oscillator-action-1} is not stable under one-loop 
corrections, and a linear counter terms needed.

It seems natural to look for a more general action. The so-called induced gauge action 
\cite{deGoursac:2007gq,Grosse:2007dm} contains the terms of \eqref{oscillator-action-1} and more. It is invariant
under noncommutative $U(1)$ transformations. The starting point is the scalar $\phi^4$ model with oscillator potential
\eqref{scalar-oscillator}. The scalar field is then coupled to an external gauge field. The dynamics of the gauge 
field is given by the divergent contributions of the one-loop effective action generalising the method of heat kernel 
expansion to the noncommutative realm. The induced action is given by
\bea
\label{induced-action}
S & = & \int d^4x\, \Bigg\{
\frac{3}{\theta} (1-\rho^2) (\tilde \mu^2-\rho^2)(\tilde X_\nu \star \tilde X^\nu -\tilde x^2) 
\\
\nonumber
&& \hspace{1.3cm}
+ \frac 32 (1-\rho^2)^2 \big( (\tilde X_\mu \star \tilde X_\mu)^{\star 2}-(\tilde x^2)^2 \big) - \frac{\rho^4}4  F_{\mu\nu} F_{\mu\nu}
\Bigg\}\,,
\eea
where $\rho = \frac{1-\Omega^2}{1+\Omega^2}$, $\tilde \mu^2 =  \frac{m^2 \theta}{1+\Omega^2}$.
Furthermore, the field strength is given by 
$$
F_{\mu\nu} = -i[\tilde x_\mu, A_\nu]_\star + i[\tilde x_\nu, A_\mu]_\star 
    -i[A_\mu,A_\nu]_\star\,,
$$ 
and $\tilde X_\mu$ denote the covariant coordinates, $\tilde X_\mu = \tilde x_\mu + A_\mu$.
In the limit $\Omega\to 0$ (i.e., $\rho\to1$), we recover the usual noncommutative Yang-Mills action. An interesting 
limit is $\Omega\to 1$ (i.e., $\rho\to 0$), where we obtain a pure matrix model. It has a non-trivial vacuum, which 
makes the quantization more difficult. The computation of propagator and Feynman rules and also one-loop calculations are work in progress.

An alternative model has been proposed in \cite{Buric:2010xs}. The gauge model is constructed on a specific curved 
noncommutative background space, the so-called truncated Heisenberg space. In two dimensions the action reads
\bea
S & = & \int d^2x\, \Big(
(1-\alpha^2) F_{12}^{*2} - 2(1-\alpha^2)\mu\, F_{12}\star \phi + (5-\alpha^2)\mu^2 \phi^2 \\
&& \hspace{1cm}
+ 4 \ri \alpha \, F_{12} \star \phi^{\star 2} + (D_i \phi)^2 - \alpha^2 \staraco{p_i + A_i}{\phi}^2 
\Big)\,,
\eea
where $\alpha$ is some parameter and $\mu$ has dimension of a mass.


\subsection{$1/p^2$ approach}

The same strategy as in \ref{pp} is applied here, the IR divergence is added as a counter term. Considering the action
\be
\label{usual-action}
S = \int d^4x \, F_{\mu\nu}\star F_{\mu\nu}
\ee 
for noncommutative $U(1)$ theory, the vacuum polarization shows the following
IR divergent contribution: 
\be
\Pi_{\mu\nu} \propto \frac {\tilde p_\mu \tilde p_\nu}{(\tilde p^2)^2}\,.
\ee
A gauge invariant implementation of the above is given by the term \cite{Blaschke:2008yj}
\be
\label{ct3}
\int d^4x\, F_{\mu\nu} \frac 1{\tilde D^2 D^2} F_{\mu\nu}\,.
\ee
The inverse covariant derivatives in the above expression need to be expanded in terms the gauge field. Hence,
vertices with arbitrary number of photon legs occur. This situation might still be treatable, but it is simpler to
use a localised version of \eqref{ct3}.
Basically, there are two different ways to implement the localization:

\begin{itemize}
\item By introducing an antisymmetric field $B_{\mu\nu}$ \cite{Blaschke:2009wf}:
\be
\int d^4x F_{\mu\nu} \frac {a^2}{\tilde D^2 D^2} F_{\mu\nu} 
\to
\int d^4x \left( a B_{\mu\nu} F_{\mu\nu} - B_{\mu\nu} \star \tilde D^2 D^2 B_{\mu\nu} \right)\,.
\ee
But this field is physical and introduces additional degrees of freedom. Therefore, the model is not pure noncommutative $U(1)$ gauge theory any more but describes different physics.

\item
Secondly, BRST doublet structures are employed in \cite{Vilar:2010zz}. The additional fields needed for the localization of \eqref{ct3} build BRST doublets. This avoids the introduction of new physical degrees of freedom. 
Unfortunately, the model presented in \cite{Vilar:2010zz} is not renormalizable.

\end{itemize}
The virtue of the latter approach is the implementation of the IR damping as a so-called "soft breaking".
This is in analogy to the Gribov-Zwanziger approach to undeformed QCD \cite{Zwanziger:1989,Zwanziger:1993}, 
where an IR modification of the propagator is suggested to cure the Gribov ambiguities. The UV renormalizability 
is not altered. In \cite{Blaschke:2009zi}, the "soft breaking" approach has been developed further. As a 
result the following action is proposed: 
\bea
\label{BRSW-action}
S & = & S_{\rm inv} + S_{\rm gf} + S_{\rm aux} + S_{\rm soft} + S_{\rm ext}\,,\\
S_{\rm inv} & = & \int d^4x\, \frac 14 F_{\mu\nu} F_{\mu\nu}\,,\\ 
S_{\rm gf} & = & \int d^4x\, s(\bar c \partial_\mu A_\mu)\,,\\
S_{\rm aux} & = &  \int d^4 x\, s \left( \bar \psi_{\mu\nu} B_{\mu\nu}\right) \,,\\
S_{\rm soft} & = & \int d^4x \, s \left(
(\bar Q_{\mu\nu\alpha\beta} B_{\mu\nu} + Q_{\mu\nu\alpha\beta} \bar B_{\mu\nu}) \frac 1{\tilde \square} (f_{\alpha\beta} + 
\sigma \frac{\theta_{\alpha\beta}}2 \tilde f)
\right)\,,\\
S_{\rm ext} & = & \int d^4x \, (\Omega_\mu^A sA_\mu + \Omega^c sc)\,,
\eea
where $f_{\alpha\beta} = \partial_\alpha A_\beta - \partial_\beta A_\alpha$ is the commutative $U(1)$ field strength, 
$\Theta_{\alpha\beta} = \epsilon\, \theta_{\alpha\beta}$ and $\tilde f = \theta_{\alpha\beta} f_{\alpha\beta}$,
$\tilde\square = \tilde\partial_\mu\tilde \partial_\mu=\theta_{\mu\alpha}\theta_{\mu\beta}\partial_\alpha\partial_\beta$. 
For convenience, $\epsilon$ has mass dimension $-2$, whereas $\theta_{\mu\nu}$ is rendered dimensionless.
The additional sources $\bar Q,Q,\bar J, J$ ensure BRST invariance of \eqref{BRSW-action}. In the IR, they take their physical values:
\bea
\nonumber
\bar Q_{\mu\nu\alpha\beta}|_{phys} & = & 0,\qquad
\bar J_{\mu\nu\alpha\beta}|_{phys} = \frac{\gamma^2}4(\delta_{\mu\alpha}\delta_{\nu\beta} - \delta_{\mu\beta}\delta_{\nu\alpha}),\\
Q_{\mu\nu\alpha\beta}|_{phys} & = & 0,\qquad
J_{\mu\nu\alpha\beta}|_{phys} = \frac{\gamma^2}4(\delta_{\mu\alpha}\delta_{\nu\beta} - \delta_{\mu\beta}\delta_{\nu\alpha})\,.
\eea
Inserting the physical values and integrating out the field $B_{\mu\nu}$ the following action is obtained:
\be
S_{\rm phys} = 
\int d^4x \left(
\frac 14 F_{\mu\nu}F_{\mu\nu} + \gamma^4 \left[\partial_\mu A_\nu
\frac 1{2 \widetilde{\square}^2} f_{\mu\nu} + \left( \sigma + \frac {\theta^2}{4} \sigma^2 \right) 
(\tilde{\partial} A)\frac 1{\widetilde{\square}^2}(\tilde{\partial} A) \right] + s \left(\bar c \partial_\mu A_\mu\right)\right)\,.
\ee
The term proportional to $\gamma^4$ breaks gauge invariance. It is called ``soft breaking'' since the parameter 
$\gamma$ has dimension of mass. We have used the commutative field strength in this expression although it is not covariant under noncommutative
gauge transformations. But it only appears in the breaking term and cannot make it worse, since gauge invariance is already violated. The advantage
is that only the propagation but not the interaction is modified due to the ``soft breaking''.

The full action \eqref{BRSW-action} is invariant under the following set of BRST transformations:
\bea
\nonumber
& sA_\mu = D_\mu c\, ,\quad sc = igcc\,,\quad s\bar c = b\,,\quad sb =0\,,\\ 
& s\bar \psi_{\mu\nu} = \bar B_{\mu\nu}\,,\quad s\bar B_{\mu\nu}=0\,,\quad
sB_{\mu\nu} = \psi_{\mu\nu}\,,\quad s\psi_{\mu\nu}=0\,,\\
\nonumber
& s\bar Q = \bar J\,,\quad s \bar J = 0\,,\quad s Q = J\,,\quad sJ=0\,.
\eea
The fields $\psi$ and $B$, resp. $\bar \psi$ and $\bar B$ and the sources $Q$ and $J$, resp. $\bar Q$ and $\bar J$ are BRST doublets.
Let us discuss the Feynman rules for \eqref{BRSW-action}. The vertex functions are the same 
as in the usual noncommutative $U(1)$ theory defined by the action \eqref{usual-action}. The propagator is more complicated, it reads
\be
G_{\mu\nu}^A(k) = \left( k^2 + \frac{\gamma^4}{\tilde k^2} \right)^{-1} \left(
\delta_{\mu\nu} - \frac{k_\mu k_\nu}{k^2} - \frac{\bar \sigma^4}{(k^2 + 
(\bar \sigma^4+ \gamma^4) \frac 1{\tilde k^2})} \frac{\tilde k_\mu \tilde k_\nu}{(\tilde k^2)^2} \right)\,,
\ee
where
$$
\bar \sigma = 2 \gamma^4 \left( \sigma + \frac{\theta^2\sigma^2}4 \right)\,.
$$
But for 1-loop calculation, it can be approximated by
\be
G_{\mu\nu}^A \sim \frac 1{k^2}(\delta_{\mu\nu} - \frac{k_\mu k_\nu}{k^2}),\quad k^2>>1\,,
\ee
since both UV and IR divergences result from high momentum range in the loop. This ignores the IR damping, but as we have seen 
the damping has no effect at one-loop. Considering higher loop insertions of a single tadpole 
(cf. \eqref{eq:canon:p2inv_gurau_n-loop-int}) the damping of the propagators between the single loops is essential 
and renders the result independent of the number of inserted loops - at least in the scalar case, for the gauge model
discussed here this still needs to be shown. 

A power counting formula,
\be
d_G = 4 - E_A - E_{c\bar c}\,,
\ee
where $E_\phi$ denotes the number of external $\phi$-legs, and one-loop results have been obtained in \cite{Blaschke:2009zi}.
The correction to the vacuum polarization is given by
\be
\Pi_{\mu\nu} = \frac{2g^2}{\epsilon^2 \pi^2} \frac{\tilde p_\mu \tilde p_\nu}{(\tilde p^2)^2} + 
\frac{13g^2}{3(4\pi)^2}(p^2\delta_{\mu\nu}- p_\mu p_\nu) \ln \Lambda\,, 
\ee
where $\Lambda$ denotes a momentum cut-off. Remarkably, the one-loop correction is transversal.
Furthermore, we obtained the following results for the vertices:
\bea
\label{eq:3-A correction_IR}
\Gamma^{\rm 3A,IR}_{\mu\nu\rho} & = & - \frac {2ig^3}{\pi^2} \cos \frac{\epsilon p_1\tilde p_2}2 
\sum_{j=1,2,3} \frac{\tilde p_{j,\mu}\tilde p_{j,\nu}\tilde p_{j,\rho}}{\epsilon (\tilde p_j^2)^2}\,,\\
\Gamma^{\rm 3A,UV}_{\mu\nu\rho} & = & - \frac{17g^2}{6(4\pi)^2} \ln \Lambda\, \tilde V^{\rm 3A,tree}_{\mu\nu\rho}(p_1,p_2,p_3)\,, \\
\Gamma^{\rm 4A,UV}_{\mu\nu\rho\sigma} & = & - \frac 5{8\pi^2} \ln \Lambda\,\, \tilde V^{\rm 4A,tree}_{\mu\nu\rho\sigma}\,,
\eea
where $V^{\rm 3A,tree}_{\mu\nu\rho}$ and $V^{\rm 4A,tree}_{\mu\nu\rho\sigma}$ denote the tree level  vertex functions.
Regarding the three-point function, the IR divergent result \eqref{eq:3-A correction_IR} corresponds to a counter term
\begin{align}
\label{3A-correction}
S^{\rm 3A,corr}=\int d^4x\, g^3\staraco{A_\mu}{A_\nu} \frac{\tilde{\partial}_\mu\tilde{\partial}_\nu\tilde{\partial}_\rho}
{\epsilon\, \tilde\square^2}A_\rho\,.
\end{align}
Such a term can readily be introduced into the ``soft breaking`` part of the action $S_{\rm soft}$ in  
\eqref{BRSW-action}. But in order to do so, we have to restore BRST invariance in the UV regime. 
Again, this can be achieved by introducing sources $Q'$ and $J'$, which form a BRST doublet,
\begin{align}
sQ' = J'\,, \qquad sJ'=0\,.
\end{align}
Consequently, we insert the following terms into $S_{\rm soft}$:
\begin{align}
\label{counter-3A}
\int d^4x \left( J'\staraco{A_\mu}{A_\nu} \frac{\tilde{\partial}_\mu \tilde{\partial}_\nu \tilde{\partial}_\rho}{\tilde \square^2}A_\rho -
Q's\left(\staraco{A_\mu}{A_\nu} \frac{\tilde{\partial}_\mu\tilde{\partial}_\nu\tilde{\partial}_\rho}{\tilde \square^2}A_\rho\right)\right)\,.
\end{align}
This term is BRST invariant by itself. In the IR, the sources take on their physical values
\begin{align}
J' = g\gamma'^2,\,\, 
Q'= 0\,,
\end{align}
and the counter term in \eqref{3A-correction} leads to a renormalization of $\gamma'$, which is another parameter of mass-dimension $1$. 

The above one-loop result leads to a negative $\beta$-function:
$$
\beta = - \frac {7g^3}{12\pi^2}\,. 
$$


\section{\label{conclusions}Concluding remarks}

The one-loop corrections for the novel action \eqref{BRSW-action} reduce to the ones known from 
the usual noncommutative $U(1)$ theory, see e.g. \cite{Matusis:2000jf,Hayakawa:1999yt}. At higher 
loop order, differences will arise. Both, UV and IR divergences can be absorbed in the tree level 
action \eqref{BRSW-action} plus \eqref{counter-3A}. But so far, a renormalization (dis)proof 
is still missing. We plan to attack this problem by applying a renormalization scheme such as 
multi-scale analysis or flow equations. \\
The negative $\beta$-function reflects the non-Abelian structure of noncommutative $U(1)$ gauge theory.

Concerning the induced gauge action \eqref{induced-action}, we plan to study the vacuum
structure, to study its quantization and as a first step to compute one-loop corrections.

\providecommand{\href}[2]{#2}\begingroup\raggedright\endgroup


\end{document}